\documentclass[aps,prl,reprint,superscriptaddress,nofootinbib]{revtex4-2}

\usepackage{amsmath,amssymb,bm}
\usepackage{microtype}
\usepackage[hidelinks]{hyperref}

\newcommand{\I}{\mathrm I}
\newcommand{\E}{\mathcal E}
\newcommand{\D}{\mathrm D}
\newcommand{\atanh}{\operatorname{arctanh}}

\begin{document}

\title{Complementary Quantum Correlations Are Universal for Qubits}

\author{Jinbo Wang}
\affiliation{School of Mathematical Sciences, Peking University,
Beijing 100871, China}
\author{Qihang Wang}
\affiliation{School of Mathematical Sciences, Peking University,
Beijing 100871, China}
\author{Kun Chen}
\email{chenkun@itp.ac.cn}
\affiliation{Institute of Theoretical Physics, Chinese Academy of Sciences,
Beijing 100190, China}
\date{August 5, 2026}

\hypersetup{
  pdftitle={Complementary Quantum Correlations Are Universal for Qubits},
  pdfauthor={Jinbo Wang, Qihang Wang, and Kun Chen}
}

\begin{abstract}
Extracting total correlations from a quantum system usually requires
reconstructing its state, whereas many experiments access only a few
measurement settings.  A possible shortcut is to add the mutual informations
obtained from complementary measurements; in dimensions above two, however,
this procedure can count the same classical correlation twice.  We establish
that qubits are protected from such overcounting.  For every two-qubit state,
the correlations observed in two complementary local bases are bounded by the
premeasurement quantum mutual information.  The proof traces this protection
to binary-entropy curvature on the Bloch ball and combines a qubit
information-exclusion tradeoff with data processing under local dephasing.
Consequently, two correlation tables give a tomography-free lower bound on
total correlation.  A score above one bit also certifies a quantitative
one-way entanglement-distillation rate; when applied to the Choi state of a
qubit channel, the same data lower bound its quantum capacity.  The theorem
therefore identifies both an operational use of complementarity and the
trusted two-dimensional setting in which its correlation accounting is valid.
\end{abstract}

\maketitle

\textit{Introduction.—}
Determining how much correlation is shared by two quantum systems is a basic
task in quantum physics.  The quantum mutual information (QMI)
$\I(A\!:\!B)=S(A)+S(B)-S(AB)$ quantifies their total classical and quantum
correlations.  Equivalently,
$\I(A\!:\!B)=\D(\rho_{AB}\Vert\rho_A\otimes\rho_B)$ is the relative-entropy
distance from the product of its marginals.  It also has an operational
interpretation as the asymptotic noise cost of erasing all correlations
\cite{Groisman2005}.  Because QMI is a nonlinear function of the joint density
operator, its direct evaluation generally requires state tomography.  Most
experiments, by contrast, record correlations in only a few spin,
polarization, or computational-basis settings.  This mismatch raises the
broadly relevant question of when limited classical readouts can certify a
guaranteed amount of correlation in the unmeasured quantum state.

For a two-level system, the natural minimal data consist of two complementary
measurements.  Two bases are mutually unbiased when a state sharp in either
basis produces uniformly random outcomes in the other
\cite{WoottersFields1989}.  Let $X$ and $Z$ denote such a pair on each
subsystem, and let $\I(X_A\!:\!X_B)$ be the classical
mutual information between the corresponding outcomes.  Schneeloch,
Broadbent, and Howell proposed the complementary-quantum correlation (CQC)
relation
\begin{equation}
 \I(X_A\!:\!X_B)+\I(Z_A\!:\!Z_B)\leq \I(A\!:\!B).
 \label{eq:cqc}
\end{equation}
If valid, Eq.~\eqref{eq:cqc} would justify adding the two observed mutual
informations without overestimating the correlation present before
measurement.  Complementarity alone does not guarantee this accounting,
because it constrains local predictability rather than whether two bipartite
tables encode the same shared classical variable.  Local unbiasedness says
that an eigenstate of $X$ is random in $Z$; it does not say that correlations
between two parties in the $X$ table and in the $Z$ table originate from
independent degrees of freedom.

This distinction also separates CQC from standard information-exclusion
relations.  Those relations limit how strongly a single memory can be
correlated with the outcomes of incompatible measurements
\cite{MaassenUffink1988,Hall1995,ColesPiani2014,Coles2017}.  In
Eq.~\eqref{eq:cqc}, by contrast,
the memories are the outcomes of different measurements on the other quantum
system.  The $X$ and $Z$ terms therefore involve two differently dephased
states, and neither a one-sided exclusion bound nor data processing alone
compares their sum with the undephased QMI.  A proof must retain quantum side
information long enough to determine how much correlation each classical
readout has discarded.

The CQC relation was proved for pure states and several structured mixed-state
families \cite{Schneeloch2014}.  Extensive numerical searches, including
fixed-purity studies of two-qubit states, found no violation
\cite{Schneeloch2014,Alsing2022}, and a recent sufficient condition covers
additional states \cite{Iqbal2026}.  The dimension dependence is nevertheless
decisive.  Companion work gives explicit counterexamples in every local
dimension $d\geq3$, where complementary readouts can count the same classical
label more than once \cite{WangCounterexamples2026}.  That construction loses
its overcounting mechanism at $d=2$.  The qubit case is therefore not merely
the smallest unresolved dimension; it tests whether binary quantum geometry
supplies a protection that complementarity lacks in general.

Here we establish Eq.~\eqref{eq:cqc} for every two-qubit density operator,
without assumptions on rank, marginals, symmetry, or separability.  The
physical origin is visible in the Bloch ball, where a qubit state is described
by three real coordinates and its entropy depends only on the Bloch radius.
We show that the curvature of this binary entropy forces information missed by
one readout to account for at least half of the information visible in the
complementary readout.  Applying this tradeoff in opposite directions on the
two parties accounts for the full crossed correlation, and data processing
under local dephasing completes the correlation budget.  This
dimension-specific mechanism yields both a sharp conceptual boundary and
practical two-setting certificates.

Full algebraic details, the boundary argument, operational corollaries, and a
proof-dependency audit are included in Appendices A--E.

\textit{Qubit half-information exclusion.—}
Let $C$ be a finite classical label encoded in an arbitrary ensemble
$\{p_c,\rho_c\}$ of qubit states.  The Holevo information
$\chi(C\!:\!Q)$ quantifies the information carried by the ensemble and bounds
the classical information accessible by measurement \cite{Holevo1973}.  For
complementary
Pauli measurements $X$ and $Z$,
\begin{equation}
 \boxed{\chi(C\!:\!Q)-\I(C\!:\!X)
 \geq\frac12\I(C\!:\!Z).}
 \label{eq:half}
\end{equation}
The label set and ranks of the conditional states are unrestricted.
Equation~\eqref{eq:half} states that the ensemble information not recovered by
the $X$ readout must account for at least half of the label information visible in
the $Z$ readout.

The state-dependent right-hand side is essential.  Conventional exclusion
bounds compare several readouts with a dimension-dependent ceiling.  Here the
quantity $\chi(C\!:\!Q)-\I(C\!:\!X)$ is instead the information retained by the
quantum carrier but lost when it is converted into the designated classical
$X$ register.  Equation~\eqref{eq:half} says that this loss is large enough to
pay half the correlation of the complementary readout.  The asymmetric form
will allow the same crossed outcome table to be paid for once from each party.

Write a qubit state as a Bloch vector $\bm v=(x,y,z)$ with $r=|\bm v|$, and
define
\begin{equation}
 h(t)=h_2\!\left(\frac{1+t}{2}\right),\qquad
 g(\bm v)=h(r)-h(x)-\frac12h(z).
 \label{eq:gdef}
\end{equation}
If $\overline{\bm v}=\sum_c p_c\bm v_c$, direct expansion gives
\begin{equation}
 \chi-\I(C\!:\!X)-\tfrac12\I(C\!:\!Z)
 =g(\overline{\bm v})-\sum_c p_cg(\bm v_c).
 \label{eq:jensen}
\end{equation}
Thus Eq.~\eqref{eq:half} is precisely Jensen's inequality if $g$ is concave
on the Bloch ball.

Replace the base-$2$ entropy by its natural-logarithm version while
differentiating; this multiplies $g$ and its Hessian by the common positive
factor $\ln 2$ and therefore does not change semidefinite signs.  At an
interior point set
\begin{equation}
 A=\frac{\atanh r}{r},\qquad C=\frac1{1-r^2},\qquad
 \bm u=\frac{\bm v}{r}.
\end{equation}
Then
\begin{align}
 -\nabla^2g&=R-D,\nonumber\\
 R&=A\mathbf 1+(C-A)\bm u\bm u^{\mathsf T},\nonumber\\
 D&=\operatorname{diag}\!\left(\frac1{1-x^2},0,
 \frac1{2(1-z^2)}\right).
 \label{eq:hessian}
\end{align}
The matrix $R$ is the full radial curvature of the qubit entropy, whereas $D$
returns the coordinate curvatures contributed by the two measured binary
distributions.  Thus concavity is reduced to showing that the radial quantum
curvature dominates these two classical directions.
Let $U=(\bm e_x,\bm e_z)$ and let $D_2$ contain the two nonzero diagonal
entries of $D$.  Since $R,D_2>0$, a Schur complement reduces
$R-UD_2U^{\mathsf T}\succeq0$ to
$Q=D_2^{-1}-U^{\mathsf T}R^{-1}U\succeq0$.  With
$X=x^2$, $Y=y^2$, $Z=z^2$, and
\begin{equation}
 \delta=\frac{1-r/\atanh r}{r^2},\qquad \delta(0)=\frac13,
\end{equation}
the reduced matrix is
\begin{equation}
 Q=\begin{pmatrix}
 \delta(Y+Z)&(1-\delta)xz\\
 (1-\delta)xz&1-Z+\delta(X+Y)
 \end{pmatrix}.
 \label{eq:Q}
\end{equation}
The scalar series
\begin{equation}
 \frac{\atanh r}{r}
 =\sum_{n\geq0}\frac{r^{2n}}{2n+1}
 \geq\sum_{n\geq0}\frac{r^{2n}}{3^n}
 =\frac1{1-r^2/3}
 \label{eq:delta}
\end{equation}
implies $\delta\geq1/3$.  Both diagonal entries of $Q$ are nonnegative, and
its determinant has the exact decomposition
\begin{align}
 \det Q={}&Z[\delta(1-Z)+(2\delta-1)X]\nonumber\\
 &+Y[\delta(1-Z)+\delta^2r^2].
 \label{eq:detQ}
\end{align}
For $\delta\geq1/2$ every term is nonnegative.  If
$1/3\leq\delta<1/2$, then $X\leq1-Z$ and the first bracket is at least
$(3\delta-1)(1-Z)\geq0$.  Hence $Q\succeq0$ and
$-\nabla^2g\succeq0$.  Continuity extends concavity to the closed Bloch ball,
so Eq.~\eqref{eq:jensen} proves Eq.~\eqref{eq:half}.  Exchanging $X$ and
$Z$ gives its axis-swapped form.

\textit{All two-qubit states.—}
Apply Eq.~\eqref{eq:half} first to Bob's ensemble conditioned on Alice's
$X$ outcome and then, with the axes exchanged, to Alice's ensemble
conditioned on Bob's $Z$ outcome:
\begin{align}
 \chi(X_A\!:\!B)-\I(X_A\!:\!X_B)
 &\geq\tfrac12\I(X_A\!:\!Z_B),\nonumber\\
 \chi(Z_B\!:\!A)-\I(Z_A\!:\!Z_B)
 &\geq\tfrac12\I(X_A\!:\!Z_B).
 \label{eq:payments}
\end{align}
Adding the two bounds therefore supplies the full crossed correlation.
The orientation is essential: the first inequality preserves Bob as a quantum
memory, whereas the second preserves Alice, so both deficits contain the same
crossed table.

Let $\E_A$ completely dephase (pinch) subsystem $A$ in the $X$ basis, and let
$\E_B$ do the same on subsystem $B$ in the $Z$ basis.  Relative-entropy data
processing
\cite{Lindblad1975} gives
\begin{equation}
 \I(A\!:\!B)+\I(X_A\!:\!Z_B)
 \geq\chi(X_A\!:\!B)+\chi(Z_B\!:\!A).
 \label{eq:cross}
\end{equation}
The two Holevo quantities are the QMI values after one local dephasing, and
$\I(X_A\!:\!Z_B)$ is the QMI after both; Eq.~\eqref{eq:cross} orders the two
correlation losses by data processing.
Indeed,
\begin{equation}
 \I(\rho)-\I(\E_A\rho)
 =\D(\rho\Vert\E_A\rho)
  -\D(\rho_A\Vert\E_A\rho_A),
 \label{eq:loss}
\end{equation}
and applying $\E_B$ to the first relative entropy yields
Eq.~\eqref{eq:cross}.

For compactness put $I_X=\I(X_A\!:\!X_B)$,
$I_Z=\I(Z_A\!:\!Z_B)$, $I_{XZ}=\I(X_A\!:\!Z_B)$,
$\chi_X=\chi(X_A\!:\!B)$, and $\chi_Z=\chi(Z_B\!:\!A)$.  Adding
Eqs.~\eqref{eq:payments} and \eqref{eq:cross} gives the exact decomposition
\begin{align}
 \I(A\!:\!B)-I_X-I_Z
={}&[\I(A\!:\!B)+I_{XZ}-\chi_X-\chi_Z]\nonumber\\
 &+[(\chi_X-I_X)+(\chi_Z-I_Z)\nonumber\\
 &\hspace{18mm}-I_{XZ}]\geq0.
 \label{eq:decomposition}
\end{align}
Every bracket is separately nonnegative.  Any qubit MUB pair is unitarily
equivalent to Pauli $X,Z$, independently on the two parties.  Singular
states follow by continuity.  This proves Eq.~\eqref{eq:cqc} for every
two-qubit density operator.  The two brackets also reveal the division of
labor in the proof: the first is an ordinary data-processing deficit, while
the second is the genuinely qubit-specific information-exclusion deficit.
The bound is tight in physically distinct regimes.  A perfectly correlated
classical bit has $(I_X,I_Z,\I)=(1,0,1)$ after choosing its encoding basis,
whereas a maximally entangled pair has $(I_X,I_Z,\I)=(1,1,2)$.

\textit{Why qubits are special.—}
The proof exposes why the dimension boundary is sharp.  A qubit density
operator has only two eigenvalues, fixed by a single Bloch radius.  This binary
spectrum gives the entropy the curvature encoded by $\delta\geq1/3$, forcing
the half-information tradeoff in Eq.~\eqref{eq:half}.  The coefficient one
half is precisely sufficient: the two orientations in
Eq.~\eqref{eq:payments} sum to the full crossed classical correlation, while
Eq.~\eqref{eq:cross} accounts for the remaining quantum information.  In
higher dimensions a density operator has more than two eigenvalues; its
entropy is therefore not controlled by a single radial coordinate.  There is
no analogous binary-curvature constraint, and mutually unbiased bases can
reveal the same encoded label in more than one setting.  The higher-dimensional counterexamples are
therefore not exceptions to a generic complementarity principle; together
with the present theorem, they show that the principle itself is specific to
binary quantum geometry \cite{WangCounterexamples2026}.

\textit{Physical implications.—}
This mechanism has direct operational consequences for estimating
correlations from limited measurements.  Define the experimentally accessible
score
\begin{equation}
 \mathcal S_{XZ}:=\I(X_A\!:\!X_B)+\I(Z_A\!:\!Z_B)
 \label{eq:score}
\end{equation}
from two $2\times2$ outcome distributions.  The theorem gives the
state-independent lower bound $\I(A\!:\!B)\geq\mathcal S_{XZ}$ for arbitrary
two-qubit sources.  It therefore replaces full state reconstruction by two
correlation tables whenever a lower bound, rather than the exact QMI, is the
relevant quantity.  No purity, Bell-diagonal structure, or source-noise model
is required for this deterministic statement.  Because QMI is a relative
entropy and an asymptotic correlation-erasure cost, the same score also
certifies that at least $\mathcal S_{XZ}$ bits per copy of total correlation
must be erased to decorrelate the state in that operational setting
\cite{Groisman2005}.

The same score quantifies distillable entanglement.  The coherent informations
$\I(A\rangle B)=\I(A\!:\!B)-S(A)$ and $\I(B\rangle A)$ are each at least
$\mathcal S_{XZ}-1$, because qubit entropies do not exceed one bit.  By the
hashing inequality, positive coherent information is an achievable asymptotic
one-way entanglement-distillation rate \cite{DevetakWinter2005}.  Hence
\begin{equation}
 D_{A\to B},D_{B\to A}\geq\max\{0,\mathcal S_{XZ}-1\},
 \label{eq:distillation}
\end{equation}
where $D$ is measured in ebits per copy under one-way local operations and
classical communication.  Thus $\mathcal S_{XZ}>1$ certifies distillability in
either direction.  The bound is tight for maximally entangled states but is not
necessary for distillability.  Since distilled ebits can be measured to
produce secret bits, the same rate is also an achievable lower bound for
one-way secret-key generation after entanglement distillation.

There is a parallel channel interpretation.  Let
$\omega_{RB}=(\mathrm{id}_R\otimes\mathcal N)(\Phi_{RA})$ be the normalized
Choi state of a qubit-input, qubit-output channel $\mathcal N$.  Its reference
marginal is maximally mixed, so the coherent information of the maximally
mixed channel input obeys
$I_c(\mathbf 1/2,\mathcal N)=\I(R\!:\!B)_\omega-1$.  Applying CQC to two
complementary measurements on $R$ and $B$, followed by the quantum coding
theorem \cite{DevetakChannel2005}, gives
\begin{equation}
 Q(\mathcal N)\geq
 \max\{0,\mathcal S_{XZ}(\omega)-1\}.
 \label{eq:channelcapacity}
\end{equation}
Thus paired measurements on transmitted Bell pairs can certify a nonzero
asymptotic quantum communication rate without reconstructing the channel.
The statement is a lower bound from one chosen channel input, not a formula
for the generally regularized capacity.

The theorem also places the two-setting data within the framework of entropic
uncertainty.  Correlation with Bob reduces Alice's outcome uncertainty by the
corresponding classical mutual information.  Combining
Eq.~\eqref{eq:cqc} with the state-dependent qubit relation
$H(X_A)+H(Z_A)\geq1+S(A)$ gives
\begin{equation}
 H(X_A|X_B)+H(Z_A|Z_B)\geq1+S(A|B).
 \label{eq:conditionalEUR}
\end{equation}
Applying the same single-system bound on both parties also yields
\begin{equation}
 H(X_AX_B)+H(Z_AZ_B)\geq2+S(AB).
 \label{eq:jointEUR}
\end{equation}
These familiar state-dependent uncertainty bounds
\cite{Berta2010,Coles2017} acquire a common interpretation here: the reduction
of uncertainty across two incompatible readouts cannot exceed the total
correlation resource in the premeasurement state.  Their left-hand sides are
obtained from the same paired outcome tables used in $\mathcal S_{XZ}$.

These operational statements have an essential scope.  They assume trusted
two-dimensional local Hilbert spaces and calibrated mutually unbiased
projective measurements.  They are not device-independent: leakage into
higher levels can restore the correlation overcounting found for $d\geq3$.
Finite-sample confidence intervals and measurement uncertainty are likewise
experimental layers beyond the deterministic inequality.  This boundary is
also informative, because it identifies Hilbert-space dimension, rather than
basis overlap alone, as the physical resource protecting the correlation
budget.

\textit{Broader implications and outlook.—}
At the level of entropy inequalities, the result is not another application
of strong subadditivity alone.  Data processing supplies
Eq.~\eqref{eq:cross}, but it cannot control the second bracket of
Eq.~\eqref{eq:decomposition}.  The missing ingredient is the concavity of the
qubit function $g$, equivalently the state-dependent half-information
exclusion inequality for arbitrary qubit ensembles.  CQC is obtained only
after two copies of this one-system entropy theorem are oriented so that their
deficits share the same crossed classical mutual information.  This
factorization isolates a reusable strategy: first identify information lost
to a local readout, then use state-space curvature to charge a complementary
correlation against that loss, and finally join the local statements by data
processing.

The dimensional boundary changes how complementarity should be used in
quantum-information certification.  The qubit theorem turns the intuition
that complementary observables are independent probes into a rigorous
correlation budget.  Higher-dimensional counterexamples reveal the missing
premise: measurement incompatibility alone does not prevent two settings from
reporting the same shared variable.  Combining their evidence therefore
requires either a dimension-specific exclusion bound or an explicit
correction for redundant correlations.

The proof suggests a route to other low-dimensional inequalities: quantify
ensemble information inaccessible to each local readout, then join oppositely
oriented bounds by data processing.  Here the factor one half is exactly what
closes the two orientations.  Which state spaces and measurement families
admit analogous closure is a concrete question beyond CQC.  Natural tests
include nonorthogonal qubit measurement axes, generalized qubit measurements,
and R\'enyi or one-shot entropies relevant to finite-blocklength protocols.
None follows automatically from the von Neumann-entropy Hessian used here.

Experimentally, the immediate target is a robust bound for approximate qubits
and imperfectly unbiased measurements, converting leakage, misalignment, and
finite-sample uncertainty into corrections to $\mathcal S_{XZ}$.  In larger
dimensions, the task is not to restore the false original relation, but to
identify dimension-dependent budgets that remain operationally measurable.

\textit{Conclusion.—}
We have proved the CQC relation for every two-qubit state and identified the
protecting mechanism as binary-entropy curvature joined with data processing.
Together with the higher-dimensional counterexamples, this establishes a
sharp boundary between binary and multilevel correlation accounting.  Within
a trusted qubit model, two complementary outcome tables therefore provide a
tomography-free lower bound on total correlation, quantitative one-way
distillation rates, and a capacity witness for qubit channels.  The broader
lesson is that complementarity alone does not make measurement records
nonredundant; that property must be established from the geometry and
dimension of the underlying quantum state space.

\begin{acknowledgments}
We thank colleagues in the CQC discussion group for helpful comments.  K.C.
acknowledges support from the Strategic Priority Research Program of the
Chinese Academy of Sciences under Grant No.~XDB1680102.  OpenAI Codex
(GPT-5.6-Sol, accessed August 2026) assisted with proof exploration,
literature organization, manuscript preparation, and executable
cross-checks.  The authors independently verified all AI-assisted results and
take full responsibility for the work.
\end{acknowledgments}

\textit{Data availability.—}
No data were created or analyzed.  All results are analytic, and executable
checks are not premises of the proof.  Focused executable cross-checks are
included in the \texttt{anc/} directory of the arXiv source archive.

\clearpage
\onecolumngrid
\appendix
These appendices give a detailed derivation of the qubit
half-information-exclusion lemma, the oriented pairing, and the
cross-pinching inequality.  It then derives the operational corollaries stated
in the Letter and records the complete proof-dependency audit.  No numerical
statement is used in any proof.

\section{Notation and reduction of qubit MUB pairs}

For a density operator $\rho$, $S(\rho)=-\operatorname{Tr}\rho\log_2\rho$.
For a bipartite quantum state,
\begin{equation}
 \I(A:B)_\rho=S(\rho_A)+S(\rho_B)-S(\rho_{AB}).
\end{equation}
For classical registers the same symbol denotes Shannon mutual information.
For an ensemble $\{p_c,\rho_c\}$ with average
$\overline\rho=\sum_c p_c\rho_c$, its Holevo information is
\begin{equation}
 \chi(C:Q)=S(\overline\rho)-\sum_c p_cS(\rho_c).
\end{equation}

Two orthonormal qubit bases are mutually unbiased if and only if their Bloch
axes are orthogonal.  A single-qubit unitary induces an arbitrary proper
rotation of the Bloch sphere.  Hence every ordered qubit MUB pair is mapped,
up to outcome relabeling and phases, to the Pauli $X,Z$ pair.  This unitary
may be chosen independently on $A$ and $B$, while all mutual informations are
unitarily invariant.  It is therefore sufficient to prove the CQC inequality
for Pauli measurements.

\section{Qubit half-information exclusion}

\subsection{Exact Jensen-gap identity}

Represent a qubit state by
\begin{equation}
 \rho(\bm v)=\frac12(\mathbf 1+x\sigma_x+y\sigma_y+z\sigma_z),
 \qquad \bm v=(x,y,z),\qquad r=|\bm v|\leq1.
\end{equation}
Define the even binary entropy function
\begin{equation}
 h(t)=-\frac{1+t}{2}\log_2\frac{1+t}{2}
      -\frac{1-t}{2}\log_2\frac{1-t}{2}.
\end{equation}
Then
\begin{equation}
 S(\rho(\bm v))=h(r),\qquad
 H(X)_{\rho(\bm v)}=h(x),\qquad
 H(Z)_{\rho(\bm v)}=h(z).
\end{equation}
For an ensemble $\{p_c,\bm v_c\}$, let
$\overline{\bm v}=\sum_c p_c\bm v_c$.  Expanding each information quantity
gives
\begin{align}
 \chi(C:Q)&=h(|\overline{\bm v}|)-\sum_cp_ch(|\bm v_c|),\\
 \I(C:X)&=h(\overline x)-\sum_cp_ch(x_c),\\
 \I(C:Z)&=h(\overline z)-\sum_cp_ch(z_c).
\end{align}
Consequently, with
\begin{equation}
 g(x,y,z)=h(\sqrt{x^2+y^2+z^2})-h(x)-\frac12h(z),
 \label{sm:eq:g}
\end{equation}
one has the exact identity
\begin{equation}
 \chi(C:Q)-\I(C:X)-\frac12\I(C:Z)
 =g(\overline{\bm v})-\sum_cp_cg(\bm v_c).
 \label{sm:eq:jensen}
\end{equation}
The information inequality in the Letter is therefore precisely Jensen's
inequality for $g$.

\subsection{Full Hessian reduction}

We prove that $g$ is concave on the closed unit ball.  Replacing the base-$2$
entropy by its natural-logarithm version multiplies $g$ and its Hessian by the
common positive factor $\ln 2$ and does not affect semidefinite signs.  The
derivatives are
\begin{equation}
 h'(t)=-\atanh t,\qquad h''(t)=-\frac1{1-t^2}.
\end{equation}
For $0<r<1$, put
\begin{equation}
 A=\frac{\atanh r}{r},\qquad C=\frac1{1-r^2},
 \qquad \bm u=\frac{\bm v}{r}.
\end{equation}
The radial Hessian formula gives
\begin{align}
 \nabla^2h(r)
 &=\frac{h'(r)}r(\mathbf 1-\bm u\bm u^{\mathsf T})
   +h''(r)\bm u\bm u^{\mathsf T}\nonumber\\
 &=-A\mathbf 1-(C-A)\bm u\bm u^{\mathsf T}.
\end{align}
The two coordinate terms in Eq.~\eqref{sm:eq:g} contribute positive diagonal
curvatures.  Hence
\begin{equation}
 -\nabla^2g=R-D,
 \quad R=A\mathbf 1+(C-A)\bm u\bm u^{\mathsf T},
 \quad D=\operatorname{diag}\!\left(
 \frac1{1-x^2},0,\frac1{2(1-z^2)}\right).
 \label{sm:eq:RD}
\end{equation}

Let $U=(\bm e_x,\bm e_z)$ and
\begin{equation}
 D_2=\operatorname{diag}\!\left(
 \frac1{1-x^2},\frac1{2(1-z^2)}\right),
 \qquad D=UD_2U^{\mathsf T}.
\end{equation}
In the open ball both $R$ and $D_2$ are positive definite.  Taking the two
Schur complements of
\begin{equation}
 \begin{pmatrix}R&U\\U^{\mathsf T}&D_2^{-1}\end{pmatrix}
\end{equation}
shows
\begin{equation}
 R-UD_2U^{\mathsf T}\succeq0
 \quad\Longleftrightarrow\quad
 D_2^{-1}-U^{\mathsf T}R^{-1}U\succeq0.
 \label{sm:eq:schur}
\end{equation}

The tangential and radial eigenvalues of $R$ are $A$ and $C$.  Therefore
\begin{align}
 R^{-1}
 &=A^{-1}\mathbf 1+(C^{-1}-A^{-1})\bm u\bm u^{\mathsf T}\nonumber\\
 &=A^{-1}\mathbf 1-(1-\delta)\bm v\bm v^{\mathsf T},
 \label{sm:eq:Rinv}
\end{align}
where
\begin{equation}
 \delta=\frac{1-A^{-1}}{r^2}
 =\frac{1-r/\atanh r}{r^2}.
 \label{sm:eq:delta}
\end{equation}
With $X=x^2$, $Y=y^2$, and $Z=z^2$, substitution gives
\begin{equation}
 Q:=D_2^{-1}-U^{\mathsf T}R^{-1}U
 =\begin{pmatrix}
 \delta(Y+Z)&(1-\delta)xz\\
 (1-\delta)xz&1-Z+\delta(X+Y)
 \end{pmatrix}.
 \label{sm:eq:Q}
\end{equation}

\subsection{Scalar certificate and determinant sign}

\par\smallskip\noindent
For $0\leq r<1$,
\begin{equation}
 A=\frac{\atanh r}{r}
 =\sum_{n=0}^{\infty}\frac{r^{2n}}{2n+1}.
\end{equation}
The elementary bound $3^n\geq2n+1$ for every integer $n\geq0$ implies
\begin{equation}
 A\geq\sum_{n=0}^{\infty}\frac{r^{2n}}{3^n}
 =\frac1{1-r^2/3}.
\end{equation}
Thus $A^{-1}\leq1-r^2/3$ and
\begin{equation}
 \delta\geq\frac13,
 \label{sm:eq:deltabound}
\end{equation}
with the continuous value $\delta(0)=1/3$.

Both diagonal entries of $Q$ are nonnegative.  Direct expansion of the
determinant, using only $r^2=X+Y+Z$, yields
\begin{align}
 \det Q
 &=\delta(Y+Z)[1-Z+\delta(X+Y)]
   -(1-\delta)^2XZ\nonumber\\
 &=Z[\delta(1-Z)+(2\delta-1)X]
   +Y[\delta(1-Z)+\delta^2r^2].
 \label{sm:eq:det}
\end{align}
If $\delta\geq1/2$, each summand in the last line is nonnegative.  If
$1/3\leq\delta<1/2$, then $2\delta-1<0$ and $X\leq1-Z$, so
\begin{equation}
 \delta(1-Z)+(2\delta-1)X
 \geq(3\delta-1)(1-Z)\geq0.
\end{equation}
Thus $\det Q\geq0$.  A real symmetric $2\times2$ matrix with nonnegative
diagonal and determinant is positive semidefinite, so $Q\succeq0$.
Equations~\eqref{sm:eq:schur} and \eqref{sm:eq:RD} give
$-\nabla^2g\succeq0$ throughout the open ball.

At $r=0$, direct differentiation gives
$-\nabla^2g=\operatorname{diag}(0,1,1/2)$.  To include the unit sphere, take
any finite set of Bloch vectors $\bm v_c$ in the closed ball, replace all of
them by $(1-\varepsilon)\bm v_c$, apply the interior Jensen inequality, and
let $\varepsilon\downarrow0$.  Binary entropy is continuous at its endpoints.
Hence $g$ is concave on the closed ball, and
Eq.~\eqref{sm:eq:jensen} proves
\begin{equation}
 \boxed{\chi(C:Q)-\I(C:X)\geq\frac12\I(C:Z)}.
\end{equation}
Interchanging $x$ and $z$ proves the axis-swapped statement.

\section{Oriented pairing and crossed local pinchings}

Let $\rho_{AB}$ be an arbitrary two-qubit state.  Measuring $A$ in $X$
prepares on $B$ an ensemble labeled by $X_A$.  Applying the lemma with
readouts $X_B,Z_B$ gives
\begin{equation}
 \chi(X_A:B)-\I(X_A:X_B)\geq\frac12\I(X_A:Z_B).
 \label{sm:eq:first}
\end{equation}
Measuring $B$ in $Z$ prepares on $A$ an ensemble labeled by $Z_B$.  Apply the
axis-swapped lemma with principal readout $Z_A$ and complementary readout
$X_A$:
\begin{equation}
 \chi(Z_B:A)-\I(Z_A:Z_B)\geq\frac12\I(X_A:Z_B).
 \label{sm:eq:second}
\end{equation}
Adding gives the oriented pairing inequality
\begin{equation}
 \chi(X_A:B)+\chi(Z_B:A)-\I(X_A:X_B)-\I(Z_A:Z_B)
 \geq\I(X_A:Z_B).
 \label{sm:eq:oriented}
\end{equation}

Let $\E_A$ be complete dephasing of $A$ in $X$ and $\E_B$ complete
dephasing of $B$ in $Z$.  They commute because they act on different
systems.  For any pinching $\E$,
\begin{equation}
 \D(\sigma\Vert\E\sigma)=S(\E\sigma)-S(\sigma).
 \label{sm:eq:pinchingD}
\end{equation}
Expanding mutual informations gives
\begin{equation}
 \I(\rho)-\I(\E_A\rho)
 =\D(\rho\Vert\E_A\rho)
  -\D(\rho_A\Vert\E_A\rho_A).
 \label{sm:eq:loss}
\end{equation}
Relative-entropy data processing under $\E_B$ yields
\begin{equation}
 \D(\rho\Vert\E_A\rho)
 \geq\D(\E_B\rho\Vert\E_A\E_B\rho).
\end{equation}
The local subtraction in Eq.~\eqref{sm:eq:loss} is unchanged because
$\E_B$ acts only on $B$.  Applying Eq.~\eqref{sm:eq:loss} again to
$\E_B\rho$ gives
\begin{equation}
 \I(\rho)-\I(\E_A\rho)
 \geq\I(\E_B\rho)-\I(\E_A\E_B\rho).
\end{equation}
The three pinched mutual informations are
\begin{equation}
 \I(\E_A\rho)=\chi(X_A:B),\quad
 \I(\E_B\rho)=\chi(Z_B:A),\quad
 \I(\E_A\E_B\rho)=\I(X_A:Z_B).
\end{equation}
Therefore
\begin{equation}
 \I(A:B)+\I(X_A:Z_B)
 \geq\chi(X_A:B)+\chi(Z_B:A).
 \label{sm:eq:cross}
\end{equation}
Adding Eqs.~\eqref{sm:eq:oriented} and \eqref{sm:eq:cross} cancels every
intermediate quantity and proves
\begin{equation}
 \boxed{\I(X_A:X_B)+\I(Z_A:Z_B)\leq\I(A:B)}.
\end{equation}
No rank, marginal, commutativity, or separability assumption on $\rho_{AB}$
was made.

\section{Operational corollaries}

Define
\begin{equation}
 \mathcal S_{XZ}=\I(X_A:X_B)+\I(Z_A:Z_B).
\end{equation}
The theorem immediately gives the state-independent lower bound
\begin{equation}
 \I(A:B)\geq\mathcal S_{XZ}.
 \label{sm:eq:qmilower}
\end{equation}
Since $\I(A:B)=\D(\rho_{AB}\Vert\rho_A\otimes\rho_B)$, this also lower-bounds
the relative-entropy distance to the product of the marginals and the
asymptotic local-randomization cost of correlation erasure \cite{Groisman2005}.

Every separable state has nonnegative conditional von Neumann entropy on both
sides.  Consequently,
\begin{equation}
 \I(A:B)=S(A)-S(A|B)=S(B)-S(B|A)
 \leq\min\{S(A),S(B)\}.
\end{equation}
For qubits $S(A),S(B)\leq1$, so Eq.~\eqref{sm:eq:qmilower} implies
\begin{equation}
 \mathcal S_{XZ}>1\quad\Longrightarrow\quad
 \rho_{AB}\ \text{is entangled}.
\end{equation}
The converse need not hold.  More strongly, define the coherent informations
\begin{align}
 \I(A\rangle B)&=-S(A|B)=\I(A:B)-S(A),\\
 \I(B\rangle A)&=-S(B|A)=\I(A:B)-S(B).
\end{align}
Because $A$ and $B$ are qubits, $S(A),S(B)\leq1$.  Combining this fact with
Eq.~\eqref{sm:eq:qmilower} yields
\begin{equation}
 \I(A\rangle B),\I(B\rangle A)\geq\mathcal S_{XZ}-1.
\end{equation}
The hashing inequality \cite{DevetakWinter2005} therefore gives the two
one-way distillation bounds
\begin{equation}
 D_{A\to B},D_{B\to A}\geq
 \max\{0,\mathcal S_{XZ}-1\}.
 \label{sm:eq:distillation}
\end{equation}
This sufficient rate is measured in ebits per input copy and is not necessary
for distillability.  Measuring the distilled ebits gives the same achievable
one-way secret-key rate.

For a qubit-input, qubit-output channel $\mathcal N$, let
\begin{equation}
 \omega_{RB}=(\mathrm{id}_R\otimes\mathcal N)(\Phi_{RA})
\end{equation}
be its normalized Choi state, with $\Phi_{RA}$ maximally entangled.  Since
$\omega_R=\mathbf1/2$,
\begin{align}
 I_c(\mathbf1/2,\mathcal N)
 &=S(B)_\omega-S(RB)_\omega\\
 &=\I(R:B)_\omega-1\\
 &\geq\mathcal S_{XZ}(\omega)-1.
\end{align}
The quantum coding theorem \cite{DevetakChannel2005} lower-bounds the
regularized quantum capacity by the coherent information of any chosen input.
Consequently,
\begin{equation}
 Q(\mathcal N)\geq
 \max\{0,\mathcal S_{XZ}(\omega)-1\}.
 \label{sm:eq:channel}
\end{equation}
This one-input achievable rate does not assert single-letter additivity.

For a qubit and complementary projective measurements, the state-dependent
entropic uncertainty relation is
\begin{equation}
 H(X_A)+H(Z_A)\geq1+S(A).
 \label{sm:eq:stateEUR}
\end{equation}
Using
$H(X_A|X_B)=H(X_A)-\I(X_A:X_B)$ and the analogous $Z$ identity,
Eqs.~\eqref{sm:eq:stateEUR} and \eqref{sm:eq:qmilower} give
\begin{align}
 H(X_A|X_B)+H(Z_A|Z_B)
 &\geq1+S(A)-\I(A:B)\nonumber\\
 &=1+S(AB)-S(B)\nonumber\\
 &=1+S(A|B).
\end{align}

Applying Eq.~\eqref{sm:eq:stateEUR} on both parties gives
\begin{align}
 &H(X_AX_B)+H(Z_AZ_B)\nonumber\\
 ={}&H(X_A)+H(X_B)+H(Z_A)+H(Z_B)-\mathcal S_{XZ}\nonumber\\
 \geq{}&2+S(A)+S(B)-\I(A:B)\nonumber\\
 ={}&2+S(AB).
\end{align}
Both corollaries have measured classical entropies on the left.  Finite-sample
confidence bounds are a separate layer beyond the deterministic theorem.

\section{Proof-dependency audit}

The logical dependencies are:
\begin{enumerate}
 \item exact entropy identities for a Bloch vector;
 \item the analytic Hessian and Schur certificate
       Eqs.~\eqref{sm:eq:RD}--\eqref{sm:eq:det};
 \item Jensen's inequality;
 \item relative-entropy data processing for two commuting local pinchings;
       and
 \item standard entropy identities for the operational corollaries.
\end{enumerate}
Random-state searches, floating-point checks, and regression tests are not
premises; removing every executable file leaves the proof unchanged.

\bibliography{references}

\end{document}